\documentstyle[sprocl,epsf]{article}

\begin{document}

\thispagestyle{empty}

\begin{flushright}
TPI-MINN-01/24 \\
UMN-TH-2008/01\\
hep-ph/0106200
\end{flushright}

\vfil

\begin{center}
{\large\bf ON GRIBOV'S IDEAS ON CONFINEMENT \footnote{
To be published in the Boris Ioffe Festschrift 'At the Frontier of Particle 
Physics / Handbook of QCD', ed. M. Shifman (World Scientific, Singapore,
2001). 
Based on the talks given 
at workshops ``Challenges in QCD'', Kfar Giladi, Israel, June 20-23,
1999 and  ``Gribov-70'', Orsay, France,  March 27-29, 2000.}}

\vspace{0.4in}

ARKADY VAINSHTEIN

\vspace{0.2in}

{\it Theoretical Physics Institute, University of Minnesota,\\
 Minneapolis, MN 55455}
\end{center}

\vfil

\begin{center}
{\bf Abstract}
\end{center}

\vspace{2mm}

I comment on possible relations of Gribov's ideas on mechanism
  of confinement with some phenomena in QCD and in supersymmetric gauge
  theories.

\newpage
\section{Gribov's mechanism of confinement of color}

V.N.~Gribov developed his ideas on confinement
in QCD over the course of more than ten years. 
While  I had a privilege of multiple discussions on the
subject  with him, 
it was difficult for me to follow. I tried to relate some of his points
to subjects I knew from QCD and from supersymmetric gauge theories.
These discussions were origin of the comments presented here. 
  
First, I will try to review briefly Gribov's ideas following his last two
papers\,\cite{qcd1,qcd2}. 
His starting point is that 
confinement is due to light quarks --- the mechanism which is similar to 
 supercritical phenomena for large $Z$ in QED. These phenomena are
 related with fermion bound states in a strong field. There is no such
 state for bosons. For this reason  Gribov believed there were no 
glueballs in pure gluodynamics, and the theory presents scaling
 behavior.  

To realize his picture of confinement in QCD Gribov introduced a number 
of new points:
\begin{itemize}
\item 
Formulation of QCD with no divergences and no need for renormalization.
Both perturbative and nonperturbative phenomena are described by Green 
functions. 
\item 
In QCD ultraviolet and
infrared regimes are
strongly interrelated to give a specific confining solution.
\item
Chiral symmetry and corresponding Goldstone bosons --- pions, play a
special role in confinement. In particular, 
there is a short distance component in the pion wave function --- a
core which is pointlike. Equations are changed because of this. 
\end{itemize}

What is the supercritical phenomenon in QED?
    A heavy nucleus with $Z>Z_{\rm cr}\sim 1/\alpha$ makes the vacuum
    of light charged fermions unstable. It forms a bound state with a
    positron at short distances (an electron component of pair goes
    away).  As a result an effective shift,  
$Z \to Z_1=Z-1$, occurs for distances larger than the bound state size. 

A similar picture is proposed for QCD based on 
    the growth of effective charge at large distances. 
    Only colorless states are stable. 
The phenomenon is described by  Gribov's equation for the fermion
propagator $G(q)$   
\begin{equation}
\label{prop}
\left(\frac{\partial}{\partial q_\mu}\right)^2 G^{-1} =
g(q)\,\frac{\partial G^{-1}}{\partial q_\mu}\, G\, \frac{\partial 
G^{-1}}{\partial q_\mu} 
-\frac{3}{16\pi^2
  f_{\pi}^2}\,\{i\gamma_5,G^{-1}\}G\{i\gamma_5,G^{-1}\}
\end{equation}
where the running coupling $g(q)$
$$
{ g(q)=\frac 4 3 \, \frac{\alpha(q)}{\pi} }
$$
comes from the solution for the gluon propagator, and the second term
is due to the pointlike structure of the pion.

\begin{figure}[h]
  \epsfxsize=7cm
  \centerline{\epsfbox{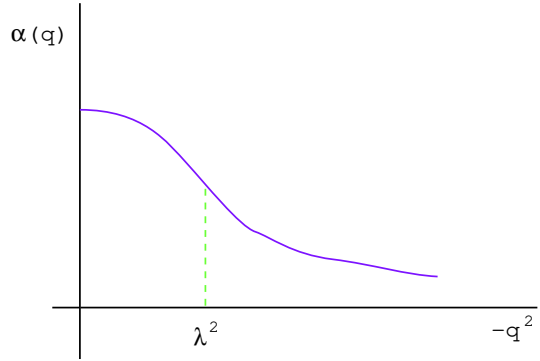}}
  \caption{Running of the effective charge $\alpha(q)$ }
  \label{fig:alpha}
\end{figure}

The quantity $\partial G^{-1}/\partial q_\mu$ in the r.h.s. of the
Gribov's equation is the effective vertex in the theory. The gluon
propagator (in Feynman gauge) is of the form
\begin{equation}
\label{gluon}
-\frac{g_{\mu\nu}}{q^2}\,4\pi \alpha(q)\;.
\end{equation}
A qualitative picture for running of the effective charge, $\alpha(q)$,
emerging from the solution for the gluon propagator is given
in Fig.~\ref{fig:alpha} 
where the scale \mbox{$\lambda \sim \Lambda_{QCD}$}.

\begin{figure}[h]
    \epsfxsize=11cm
   \centerline{\epsfbox{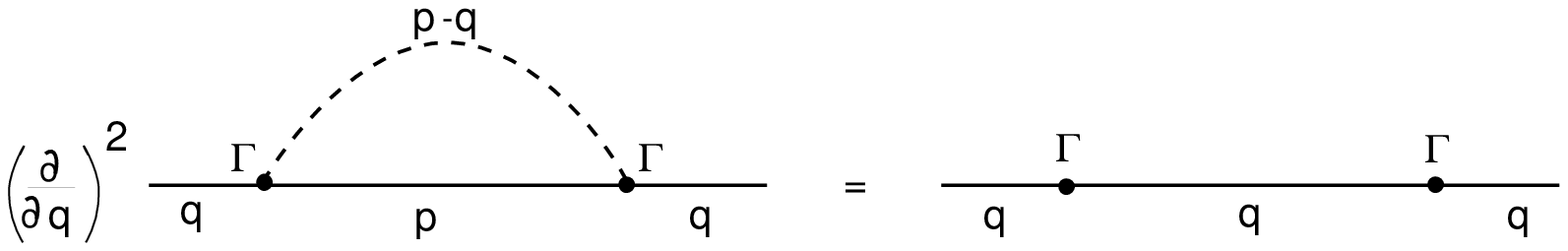}}  
    \caption{Diagrammatic form of Gribov's equation}
    \label{fig:self}
\end{figure}

Gribov's derivation of Eq.~(\ref{prop}) for the fermion propagator
is illustrated by 
Fig.~\ref{fig:self}.
Acting by $(\partial/\partial q_\mu)^2$ on the gluon propagator,
$1/(p-q)^2$, one gets $\delta^4(p-q)$. It gives the first term in
r.h.s. of Eq.~(\ref{prop}) (without the second term which is due to the
pointlike component of the pion). For   
a detail analysis of  Gribov's equation 
I refer readers to Ewerz's  work.\cite{Ewerz}

At large $q$ the solution of Eq.~(\ref{prop}) is
\begin{equation}
 G^{-1}(q) \to Z^{-1}\left[m-\not\! q + \frac{\nu_1^3}{q^2} + \frac{\nu_2^4
\not\! q }{q^4} \right]
\label{asymp}
\end{equation}
where in the limit of vanishing coupling, i.e. at  $\alpha=0$,
parameters $Z$, $m$, $\nu_1$, $\nu_2$ are
arbitrary constants.
  For perturbative  $\alpha(q)$ these parameters become running as
\begin{equation}
Z(q)=Z_0\left(\frac{\alpha(q)}{\alpha_0}\right)^{\gamma_Z},\quad
m(q)=m_0\left(\frac{\alpha(q)}{\alpha_0}\right)^{\gamma_m},\quad
\nu_{1,2}(q)=\nu^0_{1,2}\left(\frac{\alpha(q)}{\alpha_0}\right)^{\gamma_{1,2}}
\label{run}
\end{equation}
At  $m_0=0$, $\nu_1^0\neq 0$ the solution exhibits a spontaneous
chiral symmetry breaking, the coefficient $\nu_1^3$ is related to
the quark condensate, 
\begin{equation}
  \nu_1^3\propto \langle 0| \bar q q |0 \rangle\;.
\label{conden}
\end{equation}
Matching this large $q$  behavior with an absence of singularity 
at  $q\to 0$ gives what Gribov calls  a confined
  solution.
The solution features a complicated analytical structure, it leads to
an unusual filling of positive and negative energy levels.

The next step for Gribov was to calculate the meson spectrum --- he was
in the process. 

\section{SUSY gauge theories}
Supersymmetric gauge theories present an easier object than 
standard QCD. They are thoroughly studied now,\cite{SI} and one
particular  lesson
we have learned is  that the  matter content of a
theory is important for its phase. This lesson contradicts 
the ``quenched'' approach usually taken in QCD when the gluon dynamic
is presumed to play a dominant role. 

The crucial role of light matter (where light fermions
are accompanied by light bosons (pions) automatically) is explicitly seen.
Theories with SU($N_c$) gauge group and different number of flavors, $N_f$,
 produce different phases depending
on  $N_f$. In particular, supersymmetric gluodynamics ($N_f=0$) is a
confining theory,
at  $N_f=N_c - 1$ we get the Higgs phase, while in the
so called conformal window,
$$
\frac 3 2 \, N_c \le N_f \le 3N_c\;,
$$
we deal with scaling theories.

Thus, changing the number of light flavors, we are 
dramatically changing the phase of the theory. Although, the SUSY example does
not support  Gribov's  point about scaling in the case of  pure glue, it
demonstrates that his general idea about the crucial role of light
matter finds its confirmation in the SUSY world.

\section{ Duality between pion and quark propagators in OPE}

The comment below is relevant to the problem of the pointlike structure in
the pion. I am addressing here the relation between the pion and the
certain term in the OPE which was derived in.\cite{SVZ}

\begin{figure}[h]
\epsfxsize=9cm
\centerline{\epsfbox{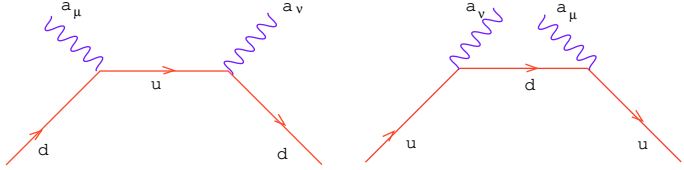}}
\caption{Diagrams for the OPE coefficients in the leading order}
\label{fig:axial}
\end{figure}

Let us consider polarization of a vacuum by the  axial current
$a_\mu=\bar u \gamma_\mu \gamma_5 d$,
\begin{equation}
 \Pi_{\mu\nu}(q)=
\int {\rm d}x\, {\rm e}^{iqx}\, \left\langle 0| T\{ a_\mu (x), \, a_\nu
  (0)\}|0 \right\rangle\;.
\label{axial}
\end{equation}
At large  $q$ we can use perturbation theory (see
Fig.~\ref{fig:axial})  to fix the
Operator  Product Expansion coefficients.
OPE allows to write $\Pi_{\mu\nu}$ as 
\begin{equation}
\Pi_{\mu\nu}=\frac{m_u+m_d}{q^4}\, q_\mu q_\nu \langle 0|\bar u u + \bar
d d|0 \rangle + \mbox{ transversal terms }
\label{axial1}
\end{equation}
(terms containing quadratic and higher powers of $m_q$ are neglected).
It is clear from Fig.~\ref{fig:axial} that the longitudinal term
(\ref{axial1}) in
the polarization operator at large $q$ is produced by linear in $m_q$
term in the quark propagator.
 
\begin{figure}[h]
\epsfxsize=4cm
\centerline{\epsfbox{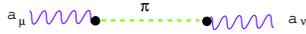}}
\caption{Pion contribution to $ \Pi_{\mu\nu}$ }
\label{fig:pion}
\end{figure}

On the other hand
we can compare this with the  
pion contribution to  $ \Pi_{\mu\nu}$, (see Fig.~\ref{fig:pion}),
which is equal to 
\begin{equation}
\Pi_{\mu\nu}^{(\pi)}= - f_\pi^2\, q_\mu q_\nu \, \frac{1}{q^2-
  m_\pi^2}\;.
\label{pion}
\end{equation}
For the longitudinal part  it gives
\begin{equation}
q^\mu\Pi_{\mu\nu}^{(\pi)}= - f_\pi^2\, q_\nu \, \frac{q^2}{q^2-
  m_\pi^2}=- f_\pi^2\, q_\nu \,\left[1+\frac{m_\pi^2}{q^2}+ 
{\cal O}(m_\pi^4)\right]\;.
\label{pionL}
\end{equation}
The leading term in this expansion in powers of $m_\pi^2$ is polynomial
in $q$, and for this reason it is not relevant. The non-polynomial pion
contribution is linear in $m_\pi^2$. It is simple to verify that 
higher states contribute to $q^\mu\Pi_{\mu\nu}$ only starting from
$m_\pi^4$ terms. 
Comparing Eq.~(\ref{pionL}) with the longitudinal part of
$\Pi_{\mu\nu}$
given by Eq.~(\ref{axial1}),  we see that
\begin{equation}
f_\pi^2 m_\pi^2 = - (m_u+m_d) \, \langle 0|\bar u u + \bar
d d|0 \rangle\;.
\label{masseq}
\end{equation}
This is a well-known relation between the quark and pion masses.

Thus, we see, that the specific term in OPE is valid all the way from large to
small $q$. At small $q$, i.e. at large distances, it is given by the
pion, but it  keeps the same form at large $q$, i.e. at short
distances, as if the pion were pointlike. 

The derivation 
of
the duality between short and long distances demonstrated above does
not imply any core in the pion. In this sense it looks as an argument
against the Gribov's idea of the pointlike structure in the pion.

\section{ Different scales in QCD}

A support for a pointlike structure in the pion comes, in my view,
from the observation of few scales in QCD, which are numerically quite
different.

To remind you of the old story, let me start with 
the low-energy theorem for the trace of energy-momentum tensor,\cite{NSVZ}
\begin{equation}
\int {\rm d}x\, {\rm e}^{iqx}\, \left\langle 0| T\left\{\frac{\alpha_s}{\pi}
 \, G_{\mu\nu}^2 (x), \frac{\alpha_s}{\pi}\, G_{\gamma\delta}^2
 (0)\right\}|0  \right\rangle
= \frac{32}{b} \left\langle 0\left| \frac{\alpha_s}{\pi}\,
    G_{\mu\nu}^2 (0)  \right|0\right\rangle\;,
\label{lowenth}
\end{equation}
where $b=11/3 \, N_c -2/3 \, N_f$. 
Numerically, the r.h.s. is quite large and leads to the large
momentum scale $\lambda^2_G$ in the correlator of two $G^2$, 
\begin{equation}
\lambda^2_G=20~ {\rm GeV}^2\;,
\label{scale}
\end{equation}
i.e. in the glueball channel with quantum numbers $0^+$. This scale
should be compared with the scale   
$m_\rho^2=0.6~ {\rm GeV}^2$ in the isovector quark channel with
quantum  numbers  $1^-$. 
The order of magnitude difference is quite significant.

\begin{figure}[h]
\epsfxsize=9cm
\centerline{\epsfbox{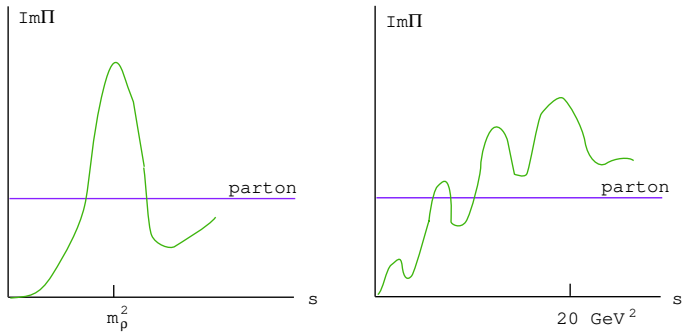}}
\caption{Duality in the $\rho$ and gluonic $0^+$  channels}
\label{fig:duality}
\end{figure}

The large scale the gluonium $0^+$ channel does not imply that the  mass
of the lowest state in this channel should be large, rather it tells
us that the onset of parton-hadron duality is at higher energies in this
channel as compared, say, with the isovector quark  $1^-$ channel.
I try to illustrate this 
 by Fig.~\ref{fig:duality}. 
In the quark channel with the $\rho$ meson
 quantum numbers the duality appears very early, it is enough to
 integrate over the $\rho$ meson peak to have it. In the gluonium
 $0^+$  channel one needs to include some number of higher excitations 
to see the duality. Numerically, $20~{\rm GeV}^2$ is the duality
interval in $s$ for the $0^+$ gluonium channel versus  $2~{\rm GeV}^2$
for the $\rho$ channel.

Gribov's view that there is no confinement in pure glue could be
related to the regime at  $E^2 < 20\, {\rm GeV}^2$.  

The existence of a much smaller spatial scale in gluon channels was
confirmed by studies in frameworks of the instanton liquid
model,\cite{shur}
and by lattice calculations.\cite{has} A particularly interesting
signal is given by the small size of the lowest $0^+$ gluonium. It
is about 0.2 $fm$ versus 1 $fm$ for a ``normal'' hadronic size.

Another channel with small spatial size is the pseudoscalar
$\bar u \gamma_5 d$ channel $0^-$,
\begin{equation}
\langle 0|\bar u \gamma_5 d|\pi\rangle \sim 
\frac{m_\pi^2}{m_q}\,f_\pi\;.
\label{gfive}
\end{equation}
Numerically, $m_\pi/m_q$ is large. As a result, the scale $f_\pi\sim$
130 MeV in the matrix element 
$\langle 0|\bar u \gamma^\mu \gamma_5 d|\pi\rangle$ of the axial
current becomes 
\begin{equation}
\frac{m_\pi^2}{m_u+m_d}\approx
2\,{\rm GeV} 
\label{gfive1}
\end{equation}
in the $\bar q \gamma_5 q$ channel.
This large scale is the key ingredient  of the penguin mechanism for 
 $\Delta \, I=1/2$ enhancement in weak decays of strange
particles (it also makes theory consistent with the recent measurement of 
 $\epsilon'/\epsilon$).\cite{peng}

I suggest that Gribov's pointlike core in the pion could be a
reflection of this large momentum scale.

\section{Concluding remarks and acknowledgments}

Gribov constructed in his mind a very physical picture of the hadronic
world which is only partially reflected in his publications. While 
some of his ideas look very unconventional I believe that they could
be the
source of a much deeper understanding of QCD than we currently have.
In my comments I limited myself to a very few examples where the depth
of Gribov's ideas can be seen.

I am thankful for discussions to Yuri Dokshitzer, Leonid Frankfurt, 
Alexey Kaidalov, Marek Karliner, Gregory Korchemsky, Al Mueller,
Mikhail Shifman, Edward Shuryak, and Mikhail Voloshin. This work
was supported by  DOE under the grant number DE-FG02-94ER408.


\section*{References}
\begin{thebibliography}{99}

\bibitem{qcd1}
V.N.~Gribov,
Eur.\ Phys.\ J.\  {\bf C10}, 71 (1999) [hep-ph/9807224].

\bibitem{qcd2}
V.N.~Gribov,
Eur.\ Phys.\ J.\  {\bf C10}, 91 (1999)
[hep-ph/9902279].

\bibitem{Ewerz}
C.~Ewerz,
Eur.\ Phys.\ J.\  {\bf C13}, 503 (2000)
[hep-ph/0001038].

\bibitem{SI}
K.~Intriligator and N.~Seiberg,
Nucl.\ Phys.\ Proc.\ Suppl.\  {\bf 45BC}, 1 (1996)
[hep-th/9509066].

\bibitem{SVZ}
M.~A.~Shifman, A.~I.~Vainshtein and V.~I.~Zakharov,
Nucl.\ Phys.\  {\bf B147}, 385 (1979).

\bibitem{NSVZ}
V.~A.~Novikov, M.~A.~Shifman, A.~I.~Vainshtein and V.~I.~Zakharov,
Nucl.\ Phys.\  {\bf B191}, 301 (1981).

\bibitem{shur}
E.~V.~Shuryak, {\it Instanton-induced effects in QCD}, this volume; 
{\it Scales and phases of non-perturbative QCD},
hep-ph/9911244;\\ 
T.~Schafer and E.~V.~Shuryak,
Rev.\ Mod.\ Phys.\  {\bf 70}, 323 (1998)
[hep-ph/9610451].

\bibitem{has}
P.~d.~Forcrand and K.~Liu,
Phys.\ Rev.\ Lett.\  {\bf 69}, 245 (1992);\\
A.~Hasenfratz and C.~Nieter,
Phys.\ Lett.\  {\bf B439}, 366 (1998)
[hep-lat/9806026].

\bibitem{peng}
A.~Vainshtein,
Int.\ J.\ Mod.\ Phys.\  {\bf A14}, 4705 (1999)
[hep-ph/9906263].

\end{thebibliography}
\end{document}